# A Two-Field-Scan Harmonic Hall Voltage Analysis For Fast, Accurate Quantification Of Spin-Orbit Torques In Magnetic Heterostructures

Xin Lin[1,2], Lijun Zhu[1,2*]

1. *State Key Laboratory of Superlattices and Microstructures, Institute of Semiconductors, Chinese Academy of Sciences, Beijing 100083, China*
2. *College of Materials Science and Opto-Electronic Technology, University of Chinese Academy of Sciences, Beijing 100049, China*

*ljzhu@semi.ac.cn

The efficiencies of the spin-orbit torques (SOTs) play a key role in the determination of the power consumption, integration density, and endurance of SOT-driven devices. Accurate and time-efficient determination of the SOT efficiencies is of great importance not only for evaluating the practical potential of SOT devices but also for developing new mechanisms for enhancing the SOT efficiencies. Here, we develop a "two-field-scan" harmonic Hall voltage (HHV) analysis that collects the second HHV as a function of a swept in-plane magnetic field at 45° and 0° relative to the excitation current. We demonstrate that this two-field-scan analysis is as accurate as the well-established but time-consuming angle-scan HHV analysis even in the presence of considerable thermoelectric effects but takes more than a factor of 7 less measurement time. We also show that the 3-parameter fit of the HHV data from a single field scan at 0°, which is commonly employed in the literature, is not reliable because the employment of too many free parameters in the fitting of the very slowly varying HHV signal allows unrealistic pseudo-solution and thus erroneous conclusion about the SOT efficiencies.

## I. Introduction

Spin-orbit torques (SOTs) are compelling in the electrical manipulation of magnetization for low-power nonvolatile magnetic memory and computing [1-5]. Since the key performances of a SOT device, including the power, scalability, and endurance, are directly related to the efficiencies of the SOTs [6-8], accurate and time-efficient quantification of the SOT efficiencies is highly preferred when the new SOT materials are considered. So far, several techniques have been developed to quantify SOT efficiencies. As discussed in Ref. [9], a reliable technique for in-plane magnetic anisotropy (IMA) samples is the "angle scan" harmonic Hall voltage (HHV) technique (Fig. 1a) [10-12]. This analysis collects the second HHV as a function of the angle ($\varphi$) of the in-plane magnetic field ($H_{xy}$) relative to the electric current under tens of different $H_{xy}$ magnitudes [9]. $H_{xy}$ is typically a few kOe to ensure a sufficiently significant HHV variation and a negligible ordinary Nernst voltage, the latter cannot be overlooked at high magnetic fields of a few Tesla [13]. When a sinusoidal electric field ($E$) is applied onto the Hall bar of an IMA macrospin sample, the out-of-phase second HHV reads [15]

$$V_{2\omega} = V_{\text{DL+ANE}} \cos\varphi + V_{\text{FL+Oe}} \cos\varphi\cos2\varphi + V_{\text{PNE}} \sin2\varphi, \quad (1)$$

with

$$V_{\text{DL+ANE}} = V_{\text{AHE}}H_{\text{DL}}/2(H_{xy}-H_{\text{k}}) + V_{\text{ANE}}, \quad (2)$$

$$V_{\text{FL+Oe}} = -V_{\text{PHE}}(H_{\text{FL}} + H_{\text{Oe}})/H_{xy}. \quad (3)$$

Here, $H_{\text{DL}}$ and $H_{\text{FL}}$ are the dampinglike and fieldlike SOT effective fields, $V_{\text{AHE}}$ the anomalous Hall voltage, $H_{\text{k}}$ the effective perpendicular anisotropy field, $V_{\text{ANE}}$ the anomalous Nernst voltage induced by the vertical thermal gradient [14], $V_{\text{PNE}}$ the planar Nernst voltage induced by the longitudinal thermal gradient (typically strong in magnetic single layers but negligible in HM/FM heterostructures [15]), and $H_{\text{Oe}}$ the transverse Oersted field exerted on the magnetic layer by the in-plane charge current. With the values of $H_{\text{DL}}$ and $H_{\text{FL}}$, the dampinglike and fieldlike SOT efficiencies can be estimated as $\xi^{j}_{\text{DL(FL)}} = (2e/\hbar)\mu_0 M_s t H_{\text{DL(FL)}}\rho_{xx}/E$ [16], where $e$ is the elementary charge, $\hbar$ the reduced Planck's constant, $\mu_0$ the permeability of vacuum, $t$ the total thickness of the magnetic layer. This analysis is very accurate but $\varphi$ scans of $V_{2\omega}$ take a long measurement time (each sample typically takes ~5 hours to measure in our case).

In this work, we develop a "two-field-scan" HHV analysis that requires only two field scans at $\varphi = 45°$ (or -45°) and $\varphi = 0°$, respectively. We show that this two-field-scan technique is as accurate as the angle-scan HHV analysis but takes an order of magnitude less time. We also show that the 3-parameter fit of the data from a single field scan at $\varphi = 0°$ typically allows unrealistic pseudo-solution.

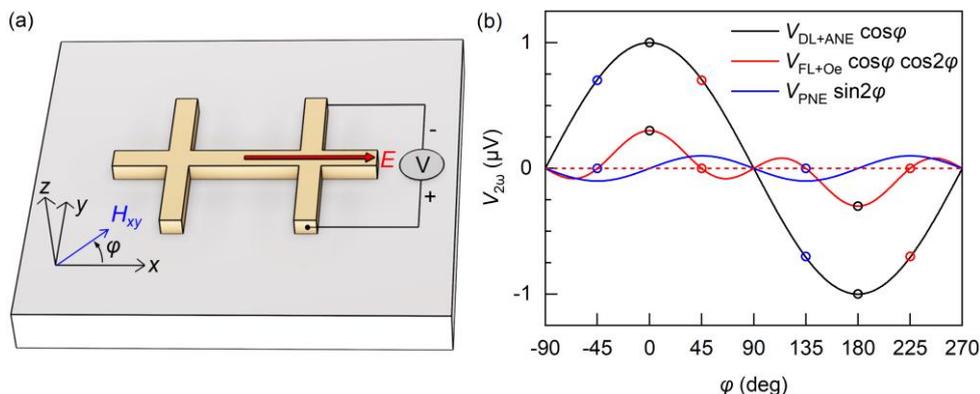

FIG. 1. (a) Schematic and coordinate of the harmonic Hall voltage (HHV) measurement. (b) Dependences of the second HHV components on the in-plane angle ($\varphi$) of the magnetic field relative to the electric field $E$.

## II. Samples and characterizations

For this study, we sputter-deposited two magnetic samples of [$Pt_{0.75}Cu_{0.25}$ (2 nm)/Co (1 nm)/Ta (1.5 nm)]$_{16}$ and $Pt_{0.75}Cu_{0.25}$ (2 nm)/Co (1 nm)/Ta (1.5 nm) with in-plane magnetic anisotropy and strong SOTs [17] on thermally-oxidized Si substrates. Each sample is seeded by a 1 nm Ta for improved adhesion and protected from oxidization by a top MgO (2 nm)/Ta (1.5 nm) bilayer. For simplicity, below we refer to the two samples as [$Pt_{0.75}Cu_{0.25}$/Co/Ta]$_{16}$ and $Pt_{0.75}Cu_{0.25}$/Co/Ta, respectively. As we show below, the [$Pt_{0.75}Cu_{0.25}$/Co/Ta]$_{16}$ exhibits a strong anomalous Nernst effect, while the $Pt_{0.75}Cu_{0.25}$/Co/Ta does not. These samples were patterned into 5 × 60 μm² Hall bars [Fig. 2(a)] by ultraviolet photolithography and argon ion milling. The resistivity $\rho_{xx}$ of the 2 nm $Pt_{0.75}Cu_{0.25}$ is 76 μΩ cm for the $Pt_{0.75}Cu_{0.25}$/Co/Ta and 93 μΩ cm for the [$Pt_{0.75}Cu_{0.25}$/Co/Ta]$_{16}$. The transport measurements are performed under a vector magnetic field that is monitored by a three-axis magnetic field sensor.

As shown in Fig. 2(b), both the [$Pt_{0.75}Cu_{0.25}$/Co/Ta]$_{16}$ and the $Pt_{0.75}Cu_{0.25}$/Co/Ta exhibit a quick magnetization increase in the small field region and saturation magnetization (1500 ± 8 emu/cm³ for the $Pt_{0.75}Cu_{0.25}$/Co/Ta, and 1520 ± 7 emu/cm³ for the [$Pt_{0.75}Cu_{0.25}$/Co/Ta]$_{16}$) when $H_{xy}$ is greater than the in-plane saturation field $H_{s,in}$ of 0.65 kOe. For the HHV measurements, we apply a sinusoidal electric field of $E$ = 46.00 kV/m via a lock-in amplifier onto the Hall bar devices along the $x$ direction (Fig. 1(a)). $V_{AHE}$ and $H_k$ are extracted from the dependence of in-phase first HHV ($V_{1\omega}$) on the swept out-of-plane magnetic field ($H_z$) under zero $H_{xy}$ [Fig. 2(c)], while $V_{PHE}$ is determined by the fit of the $\varphi$ dependence of $V_{1\omega}$ under zero $H_z$ to the relation $V_{1\omega} = V_{PHE} \sin 2\varphi$.

The "true" values of the effective SOT fields ($H_{DL}$ and $H_{FL}$) and $V_{ANE}$ of the samples are characterized using the well-established but time-consuming "angle scan" HHV analysis [9-11] with $H_{xy}$ in the region of 0.25-4.0 kOe. In Fig. 2(d) we show the representative $V_{2\omega}$ data as a function of $\varphi$ at $H_{xy}$ = 3.75 kOe for the [$Pt_{0.75}Cu_{0.25}$/Co/Ta]$_{16}$ and for the $Pt_{0.75}Cu_{0.25}$/Co/Ta. The fits of the $\varphi$ dependences of $V_{2\omega}$ to Eq. (1) yield the values of $V_{DL+ANE}$, $V_{FL+Oe}$, and $V_{PNE}$ for different $H_{xy}$. Interestingly, $V_{2\omega}$ of the [$Pt_{0.75}Cu_{0.25}$/Co/Ta]$_{16}$ is of the opposite sign compared to that of the $Pt_{0.75}Cu_{0.25}$/Co/Ta because the anomalous Nernst effect in this thick [$Pt_{0.75}Cu_{0.25}$/Co/Ta]$_{16}$ is so strong such that $V_{ANE}$ dominates the variation of the $V_{2\omega}$ signal [see Eq. (2)].

As shown in Fig. 2(e), the $V_{DL+ANE}$ data is a good linear function of $V_{AHE}/2/(H_{xy} + H_k)$ in the field region of $H_{xy} \geq H_{s,in}$ for both samples. According to Eq. (2), the slope and the intercept of the linear fit of $V_{DL+ANE}$ vs $V_{AHE}/2/(H_{xy} - H_k)$ give the values of $H_{DL}$ and $V_{ANE}$, respectively. However, the data points from the small fields below the in-plane saturation field $H_{s,in}$ ($H_{xy}$ = 0.25 and 0.5 kOe), which are blue-marked in Fig. 2(e), deviate from the linear scaling of Eq. (2) due to the unsaturation (see the hysteresis loops in Fig. 2(b)). According to Eq. (3), the slope of the linear fit of $V_{FL+Oe}$ vs $-V_{PHE}/H_{xy}$ is the sum value $H_{FL} + H_{Oe}$, from which $H_{FL}$ [Fig. 2(f)] is determined after subtraction of the Oersted field $H_{Oe}$ (0.62 Oe for the [$Pt_{0.75}Cu_{0.25}$/Co/Ta]$_{16}$ and 0.76 Oe for the $Pt_{0.75}Cu_{0.25}$/Co/Ta [17]). Thus, from the angle-scan HHV analysis, "true" values of the SOT fields and anomalous Nernst voltages are characterized as $H_{DL}$=14.2 ± 0.3 Oe, $H_{FL}$=-1.63 ± 0.1 Oe, and $V_{ANE}$ = -7.9 ± 0.1 μV for the [$Pt_{0.75}Cu_{0.25}$/Co/Ta]$_{16}$; $H_{DL}$=17.5 ± 0.3 Oe, $H_{FL}$=-0.93 ± 0.05 Oe, and $V_{ANE}$ = -0.62 ± 0.06 μV for the $Pt_{0.75}Cu_{0.25}$/Co/Ta.

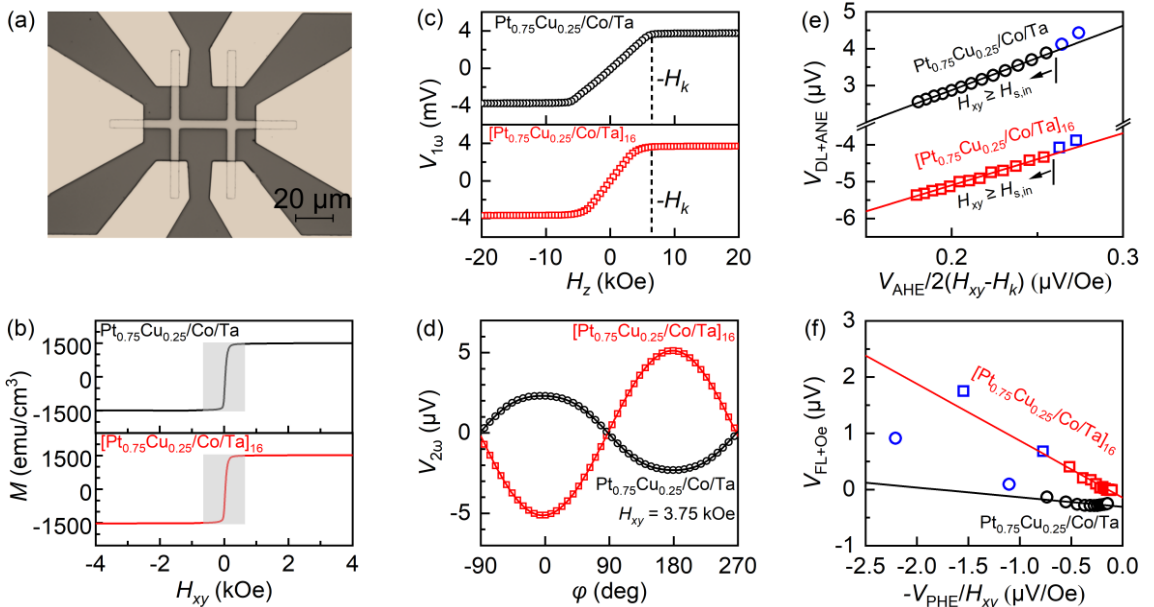

FIG. 2. Characterizations of the [$Pt_{0.75}Cu_{0.25}$/Co/Ta]$_{16}$ and the $Pt_{0.75}Cu_{0.25}$/Co/Ta. (a) Optical microscopy image of a typical Hall bar device, (b) In-plane magnetization plotted as a function of the in-plane magnetic field, with the gray area highlighting the gradual saturation of magnetization at magnetic fields below the in-plane saturation field ($H_{xy} \leq 0.65$ kOe). (c) First HHV ($V_{1\omega}$) vs the out-of-plane magnetic field ($H_z$), (d) Second HHV ($V_{2\omega}$) vs $\varphi$ ($H_{xy}$ = 3.75 kOe), (e) $V_{DL+ANE}$ vs $V_{AHE}/2/(H_{xy} - H_k)$, and (f) $V_{FL+Oe}$ vs $-V_{PHE}/H_{xy}$ for the [$Pt_{0.75}Cu_{0.25}$/Co/Ta]$_{16}$ and the $Pt_{0.75}Cu_{0.25}$/Co/Ta. The solid curve in (d) represents the fit of the data to Eq. (1), and the straight lines in (e) and (f) represent the best linear fits of the data measured under in-plane fields greater than the saturation field (0.65 kOe). In (e) and (f) the blue data points measured at magnetic fields below the saturation field deviate from the linear scaling predicted from Eqs. (2) and (3).

## III. Two-field-scan HHV analysis.
### III.A. Modeling of the two-field-scan HHV analysis

Since the "angle-scan" HHV analysis is very time-consuming and requires a magnet continuously rotatable relative to the sample, we propose a very fast but accurate analysis scheme which we term as "two-field-scan HHV analysis". For the sack of simplicity of field scan analysis, we redefine the angle and field as

$$\varphi_0 \equiv \begin{cases} \varphi, & \text{for } \varphi \leq 90°, \\ \varphi - 180°, & \text{for } \varphi > 90° \end{cases}$$

$$H_{xy0} \equiv \begin{cases} H_{xy}, & \text{for } \varphi \leq 90° \\ -H_{xy}, & \text{for } \varphi > 90° \end{cases}$$

According to Eqs. (1)-(3), the second HHV at $\varphi_0 = \pm 45°$ and 0° reduces to Eqs. (4) and (5), i.e.,

$$V_{2\omega}(\varphi_0 = \pm 45°) = \text{sign}(H_{xy0})[\sqrt{2} V_{AHE} H_{DL}/4(|H_{xy0}|-H_k) + \sqrt{2} V_{ANE}/2] \pm V_{PNE}, \quad (4)$$

$$V_{2\omega}(\varphi_0 = 0°) = \text{sign}(H_{xy0})[V_{AHE} H_{DL}/2(|H_{xy0}|-H_k) + V_{ANE} - V_{PHE}(H_{FL} + H_{Oe})/|H_{xy0}|]. \quad (5)$$

In Eq. (4) only the term containing $H_{DL}$ is dependent on both the magnitude and the sign of $H_{xy0}$, the $V_{ANE}$ term is only dependent on the sign of $H_{xy0}$, $V_{PNE}$ is the field-independent offset. which allows the reliable determination of $H_{DL}$, $V_{ANE}$, and $V_{PNE}$ from a three-parameter fit of the $V_{2\omega}(\varphi_0 = \pm 45°)$ data vs $H_{xy0}$ to Eq. (4). With the obtained value of $H_{DL}$ and $V_{ANE}$, the one-parameter non-linear fit of the $V_{2\omega}(\varphi_0 = 0°)$ data to Eq. (5) yields the values of $H_{FL}$. So far, all the parameters (i.e., $H_{DL}$, $H_{FL}$, $V_{ANE}$, and $V_{PNE}$) would be determined from the two-field-scan HHV analysis, with one field scan at $\varphi_0 = +45°$ (or -45°) and another at $\varphi_0 = 0°$. Experimentally, the value of $V_{PNE}$ should also be corrected if the lock-in amplifier has a non-zero, $\varphi_0$-independent offset signal as can be determined from the $V_{2\omega}$ value at $\varphi_0 = 90°$ (Appendix A1).

### III.B. Robustness of the two-field-scan HHV analysis

We first verify the accuracy and robustness of the two-field-scan HHV analysis using the [Pt$_{0.75}$Cu$_{0.25}$/Co/Ta]$_{16}$ with strong anomalous Nernst effect. As shown in Figs. 3(a)-3(c), the $V_{2\omega}$ data of the [Pt$_{0.75}$Cu$_{0.25}$/Co/Ta]$_{16}$ is collected as a function of the swept $H_{xy0}$ at $\varphi_0 = \pm 45°$ and 0° and fitted to Eqs. (4) and (5), respectively. The values of $H_{DL}$, $H_{FL}$, $V_{ANE}$, and $V_{PNE}$ estimated from two-field-scan HHV analysis of the data in the field region with a magnitude greater than $H_0 = 650$ Oe coincide very well with the "true" values determined from the angle-scan HHV analysis (see Figs. 3(d)-3(e)), which verifies the accuracy of the newly developed two-field-scan HHV analysis.

We then test the robustness of the two-field-scan HHV analysis against the misalignment of the sample from the intended $\varphi_0$ angles (e.g., ±45° and 0°). This evaluation is necessary since a small misalignment $\Delta\varphi_0$ can easily arise from the uncertainty of the sample mounting or the lithography misalignment between the Hall bar and the device electrodes. For this purpose, the $V_{2\omega}$ data of the [Pt$_{0.75}$Cu$_{0.25}$/Co/Ta]$_{16}$ is collected as a function of the swept $H_{xy0}$ at different $\varphi_0$ values of ±45°-$\Delta\varphi_0$ and 0°-$\Delta\varphi_0$ (-12.5° ≤ $\Delta\varphi_0$ ≤ 12.5°) and fitted to Eqs. (4) and (5), respectively (see Figs. 3(a)-3(c) for the representative results of the [Pt$_{0.75}$Cu$_{0.25}$/Co/Ta]$_{16}$). As shown in Figs. 3(d) and 3(e), the estimated values of $H_{DL}$ and $V_{ANE}$ increasingly deviate from the true values as the misalignment |$\Delta\varphi_0$| increases. The relative deviations are ≤ 5% at |$\Delta\varphi$| ≤ 2° and ≈26-30% at |$\Delta\varphi$| =12.5°. As shown in Fig. 3(f), $H_{FL}$ shows strong robustness against misalignment and coincides well with that from the angle-scan HHV analysis in the wide range of |$\Delta\varphi$| ≤ 7.5°. Therefore, the two-field-scan HHV analysis allows accurate determination of $H_{DL}$, $H_{FL}$, and $V_{ANE}$ when the angle-misalignment is not greater than 2° which can be easily ensured in experiments.

Next we show that it is critical to choose a good starting magnetic field $H_0$ for the fit analysis. In Figs. 3(g)-3(i), we summarize the values of $H_{DL}$, $V_{ANE}$, and $H_{FL}$ of the [Pt$_{0.75}$Cu$_{0.25}$/Co/Ta]$_{16}$ estimated at $\Delta\varphi_0 = 0°$ as a function of $H_0$. When the starting field $H_0$ is smaller than the in-plane saturation field $H_{s,in}$ of 0.65 Oe, $H_{DL}$ is overestimated while $V_{ANE}$ and $H_{FL}$ are underestimated. This is attributed to the breakdown of macrospin approximation at very low fields (see the low-field region of the hysteresis loop in Fig. 2b). When a $H_0$ is much greater than the saturation field [Fig. 3(i)] is chosen such that only a narrow field region is used for the fit, $H_{DL}$ becomes underestimated but $V_{ANE}$ and $H_{FL}$ are overestimated. This is because the second HHV contributions of the dampinglike torque and the fieldlike torque are inversely dependent on $H_{xy0}$ and vary more rapidly at small fields but slowly at high fields. As shown in Figs. 4(a)-4(i), the same conclusions can be made for the Pt$_{0.75}$Cu$_{0.25}$/Co/Ta. These observations suggest that it is critical to make full use of the available macrospin data ($H_0 = H_{s,in}$) for the accurate determination of SOTs and the ANE from the two-field-scan HHV analysis. In the above discussion, we have ignored the influence of the ordinary Nernst voltage because it is typically negligible at magnetic fields of a few thousand kOe [13].

## IV. Inaccuracy of the single-field-scan HHV analysis

The SOTs of the samples were also characterized in the literature using the "single field scan" HHV technique [18-23], which collects $V_{2\omega}$ from a single scan of the in-plane magnetic field along the current direction ($H_x$, $\varphi_0 = 0°$) and then estimates $H_{DL}$, $H_{FL}$, and $V_{ANE}$ from a 3-parameter nonlinear fit of the data to Eq. (7) [Fig. 5(a)]. As shown in Fig. 5(b), both $H_{DL}$ is substantially underestimated (by 33% in the case of the [Pt$_{0.75}$Cu$_{0.25}$/Co/Ta]$_{16}$) but $H_{FL}$ and $V_{ANE}$ overestimated (e.g., $H_{FL}$ is estimated by a factor of 2 for the [Pt$_{0.75}$Cu$_{0.25}$/Co/Ta]$_{16}$), regardless of the magnitude of $\Delta\varphi_0$. The inaccuracy of the 3-parameter fit of the single field scan arises because the second HHVs of the different contributions have very weak but similar dependences on the in-plane fields such that the mathematically best fit is not necessarily the correct fit. The same conclusions are true for the Pt$_{0.75}$Cu$_{0.25}$/Co/Ta sample as can be seen from Figs. 5(c) and 5(d). In the literature [18-24], this analysis is also employed to measure the SOTs of the PMA heterostructures that are aligned in-plane by very high in-plane magnetic fields. However, in that case care is needed because the ordinary Nernst voltage may not be overlooked at high magnetic fields of a few Teslas [13].

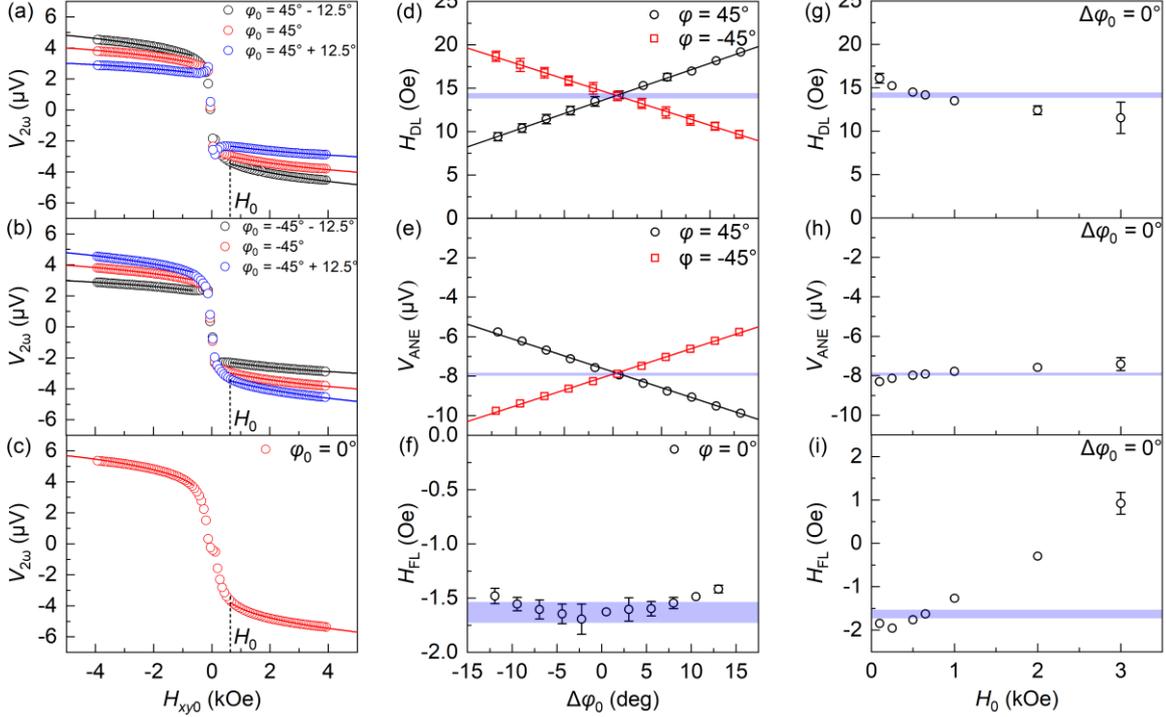

FIG. 3 Dependence on the in-plane magnetic field of the second HHV of the $[Pt_{0.75}Cu_{0.25}/Co/Ta]_{16}$ measured at (a) $\varphi_0=45°$, $45°\pm12.5°$, (b) = $-45°$, $-45°\pm12.5°$, and (c) $\varphi_0 = 0°$. The solid curves in (a)-(c) plot the best fits of the data to Eqs. (4) and (5), respectively. Influence of the misalignment $\Delta\varphi_0$ on (d) $H_{DL}$, (e) $V_{ANE}$, and (f) $H_{FL}$ as estimated for the fits of the data points of the $[Pt_{0.75}Cu_{0.25}/Co/Ta]_{16}$ in the field range of $\geq 0.65$ kOe ($H_0 = 0.65$ kOe). The solid lines in (d) and (e) are to guide eyes. Influence of the lower-bound field ($H_0$) on (g) $H_{DL}$, (h) $V_{ANE}$, and (i) $H_{FL}$ as estimated for the fits of the data points of the $[Pt_{0.75}Cu_{0.25}/Co/Ta]_{16}$ to Eqs. (4) and (5). The blue band in (d)-(i) represents the true values of $H_{DL}$, $V_{ANE}$, and $H_{FL} + H_{Oe}$ as measured from the angle scan HHV analysis with the width representing the error.

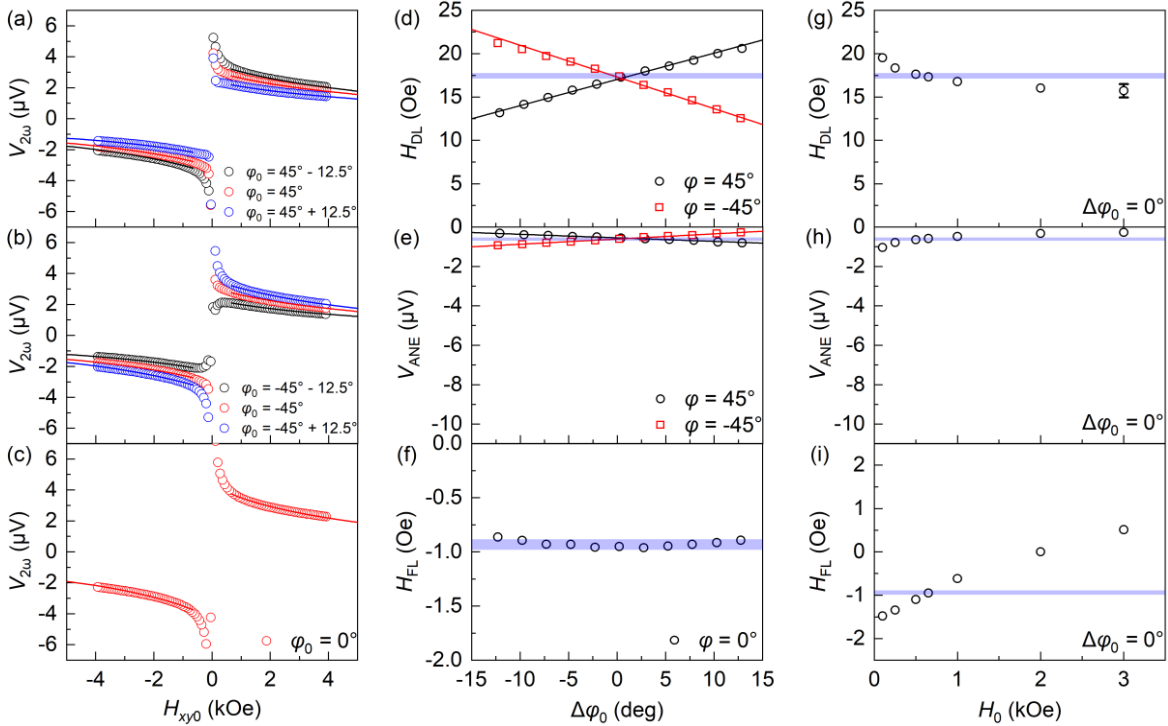

FIG. 4. Dependence on the in-plane magnetic field of the second HHV of the $Pt_{0.75}Cu_{0.25}/Co/Ta$ measured at (a) $\varphi_0=45°$, $45°\pm12.5°$, (b) $\varphi_0= -45°$, $-45°\pm12.5°$, and (c) $\varphi_0 = 0°$. The solid curves in (a)-(c) plot the best fits of the data to Eqs. (4) and (5), respectively. Influence of the misalignment $\Delta\varphi_0$ on the results of (d) $H_{DL}$, (e) $V_{ANE}$, and (f) $H_{FL}$ as estimated for the fits of the data points of the $Pt_{0.75}Cu_{0.25}/Co/Ta$ in the field range of $\geq 0.65$ kOe ($H_0 = 0.65$ kOe). The solid lines in (d) and (e) are to guide eyes. Influence of the starting magnetic field ($H_0$) on the results of (g) $H_{DL}$, (h) $V_{ANE}$, and (i) $H_{FL}$ as estimated for the fits of the data points of the $Pt_{0.75}Cu_{0.25}/Co/Ta$ to Eqs. (4) and (5). The blue band in (d)-(i) represents the true values of $H_{DL}$, $V_{ANE}$, and $H_{FL}$ as measured from angle-scan HHV analysis.

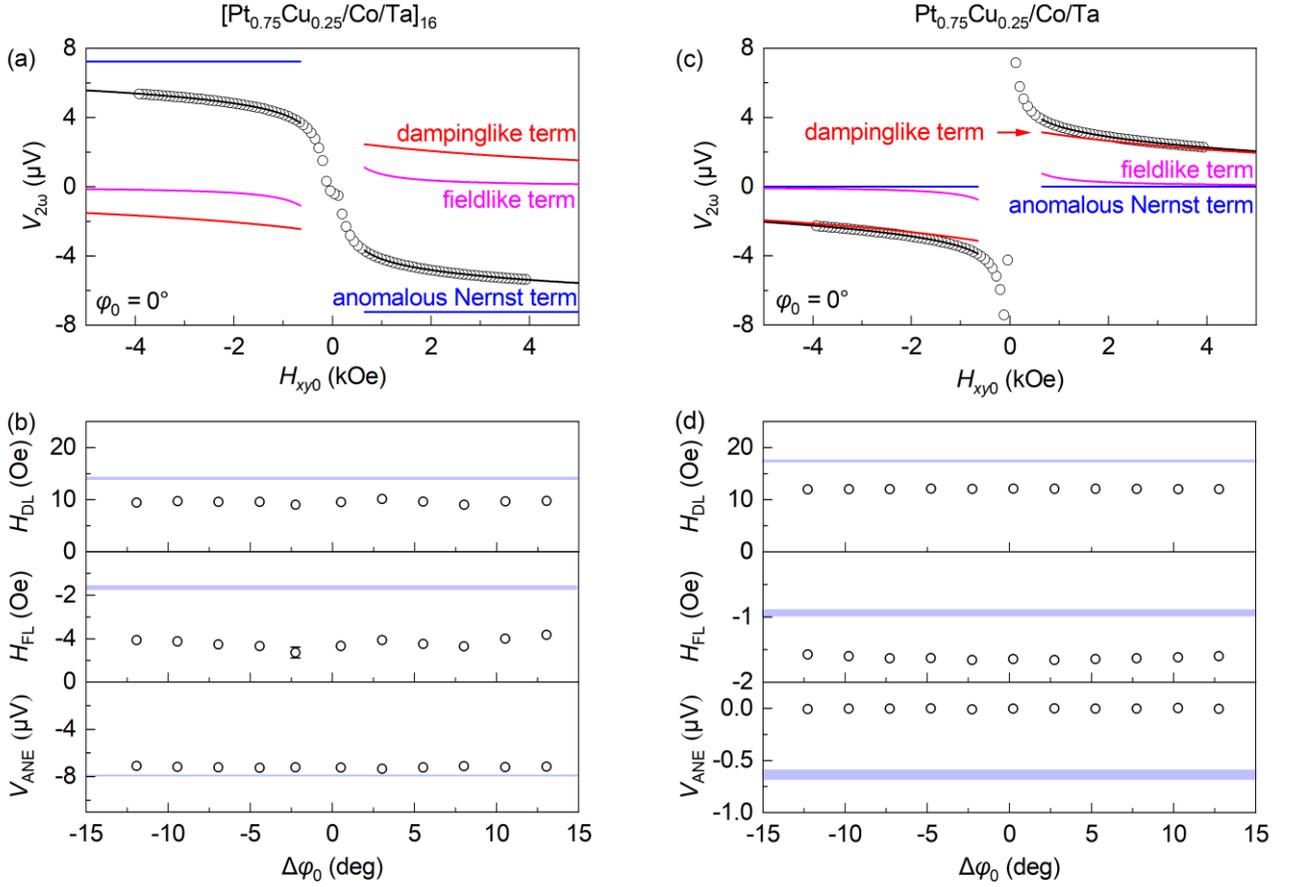

FIG. 5. Single-field-scan HHV analysis of the $[Pt_{0.75}Cu_{0.25}/Co/Ta]_{16}$ and the $Pt_{0.75}Cu_{0.25}/Co/Ta$. (a) $V_{2\omega}$ vs $H_{xy0}$ ($\varphi_0 = 0°$) for the $[Pt_{0.75}Cu_{0.25}/Co/Ta]_{16}$. (b) Dependences on $\Delta\varphi_0$ of the values of $H_{DL}$, $H_{FL}$, and $V_{ANE}$ of the $[Pt_{0.75}Cu_{0.25}/Co/Ta]_{16}$ as extracted from the fits of the data to Eq.(5) in (a). (c) $V_{2\omega}$ vs $H_{xy0}$ ($\varphi_0 = 0°$) for the $Pt_{0.75}Cu_{0.25}/Co/Ta$. (d) Dependences on $\Delta\varphi_0$ of the values of $H_{DL}$, $H_{FL}$, and $V_{ANE}$ of the $Pt_{0.75}Cu_{0.25}/Co/Ta$ as extracted from the fits of the data to Eq.(5) in (c). In (a) and (c) the black curves represent the best fits of the data to Eq.(5) with the three free parameters $H_{DL}$, $V_{ANE}$, and $H_{FL}$; the red, pink, and blue curves represent the three terms of the $V_{2\omega}$ signals, i.e., the dampinglike torque term of $sign(H_{xy0})V_{AHE}H_{DL}/2(|H_{xy0}| - H_k)$, the fieldlike torque term of $- sign(H_{xy0})V_{PHE}(H_{FL} + H_{Oe})/|H_{xy0}|$, and the anomalous Nernst term of $sign(H_{xy0})V_{ANE}$, respectively. The blue bands in (b) and (d) represent the values of $H_{DL}$, $H_{FL}$, and $V_{ANE}$ determined from angle-scan HHV analyses.

In summary, we have developed a "two-field-scan" harmonic Hall voltage (HHV) analysis that requires only two magnetic field scans at in-plane angles $\varphi_0 = 45°$ (or -45°) and $\varphi_0 = 0°$, respectively. We find that this two-field-scan analysis is as accurate as the well-established angle-scan HHV analysis but takes more than a factor of 7 less time. The two-field-scan technique is experimentally friendly since it does not require a rotational magnetic field. In contrast, the 3-parameter nonlinear fit of the HHV data from a single field scan at $\varphi_0 = 0°$ [18-23] is not reliable because the employment of too many free parameters in the fitting of the very slowly varying HHV signal yields unrealistic pseudo-solution and thus erroneous conclusion about the SOT efficiencies.

## ACKNOWLEDGMENTS

This work is supported partly by the National Key Research and Development Program of China (2022YFA1204004), by the Beijing Natural Science Foundation (Z230006), partly by the National Natural Science Foundation of China (12274405, 12304155), and partly by the Strategic Priority Research Program of the Chinese Academy of Sciences (XDB44000000).

**Appendix**

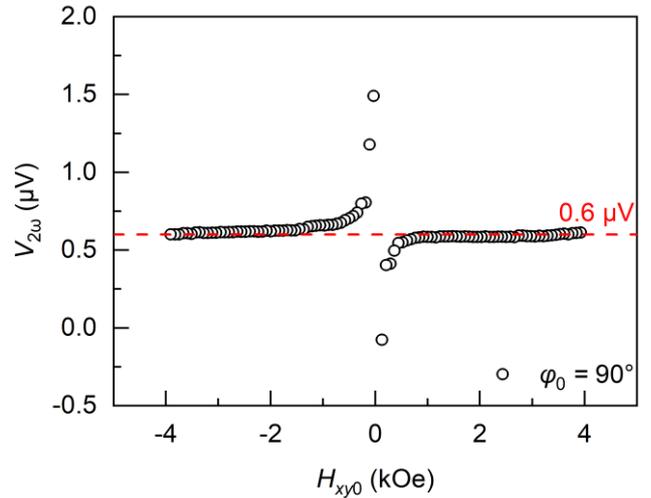

**Fig. A1.** $V_{2\omega}$ vs $H_{xy0}$ ($\varphi_0 = 90°$) of the $[Pt_{0.75}Cu_{0.25}/Co/Ta]_{16}$. The dashed line indicates a small background instrument signal from the lock-in amplifier. This instrument signal is independent of the sign and magnitude of in-plane magnetic field when the sample is saturated by the in-plane magnetic field. The sample should have no $V_{2\omega}$ signal at $\varphi_0 = 90°$, see Eq. (1).